# Aspects of High Energy Nuclear Physics : The Higgs[1]

B. Ghosh and S. Aziz
*Department of Physics, The University of Burdwan, Burdwan - 713104*

**Abstract:** Although the Higgs particle is at the centre-stage of the Standard Model of particle physics and in many other related models, it has not been experimentally found so far. One of the major objectives of the Large Hadron Collider experiments is to detect the Higgs. In this article, we highlight the problems encountered in the Higgs sector of particle physics and their possible solutions, with particular emphasis on the Little Higgs Model. We also indicate the reaction channels by which the Higgs can possibly be found in high energy accelerator experiments.

## 1. Introduction

The origin of the Higgs physics can be linked back to the quest for the origin of the masses of the weak gauge bosons. The very short range of the weak interaction entails large masses of the weak gauge bosons as per the *uncertainty principle*. A gauge boson mass term in the Lagrangian, however, violates local gauge invariance. Then it was found that the gauge bosons can be given mass in a gauged scalar field theory by a phenomenon called *spontaneous symmetry breaking* (SSB), which does not disturb the local gauge symmetry. This scalar field gets mass by SSB and the gauge bosons which interact with it also acquire masses. The scalar field under discussion is called the *Higgs field*, after Peter Higgs [1] who introduced a unitary gauge to do away with the unwanted and unobserved massless Goldstone bosons that are necessarily associated with the spontaneous breaking of a continuous symmetry, according to the famous *Goldstone Theorem*[2].

[1]Invited talk delivered by Prof. B. Ghosh at the National Seminar on '*Nuclear Physics Research in India-Facilities and Perspectives'*, held at Ranigunge Girls' College' Ranigunge, Burdwan, on February 25-26, 2010.

## 2. The quantum correction to the Higgs mass

In quantum field theory, the quantum correction to a physical quantity is obtained by evaluating perturbative loop diagrams. In calculating the amplitudes from such diagrams, we have to integrate over the loop momentum. Such integrals may diverge sometimes; then one has to see how to tackle such divergences. Let us consider the case of quantum correction to the Higgs mass in the SM, which may result from its interaction with a gauge boson or a fermion.

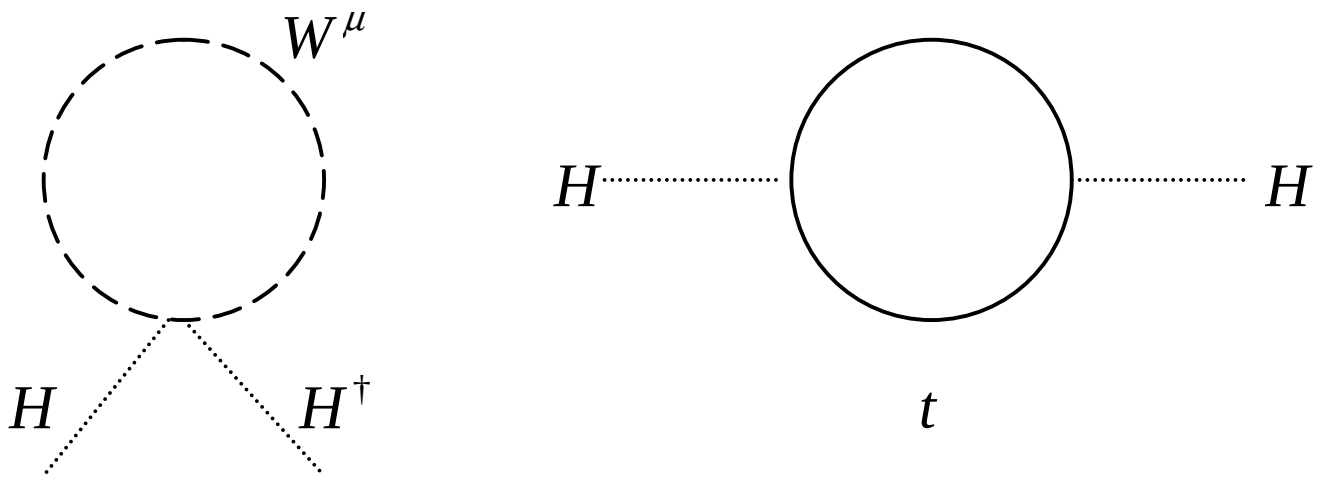

($a$) ($b$)

Fig.1. One-loop diagrams contributing to the quantum correction of the Higgs mass in the SM: (a) gauge-boson loop, (b) top quark loop.

The quantum corrections to (*Higgs mass*)$^2$ due to Figs. 1(a) and 1(b) respectively are, $\delta m_h^2 \approx g^2\Lambda^2/16\pi^2$ and $\delta m_h^2 \approx Y_t^2\Lambda^2/16\pi^2$, where $\Lambda$ is a cut-off value of the loop momentum and $g, Y_t$ are coupling constants. Obviously, the Higgs mass is unstable in the SM. This is sometimes called, the '*little hierarchy problem*' [3]. Experiments [4] have suggested a lower bound of the SM Higgs mass as, $m_h \geq 114.4$ GeV. Assuming, $\delta m_h^2 \approx m_h^2 \simeq 115$ GeV, we get $\Lambda \simeq 2.3$ TeV. Thus 'new physics' must start at around this energy. The new physics is described by models which are '*beyond Standard Model*' (BSM).

**3. Higgs in BSM**

There are several BSM's, where Higgs has been interpreted and described in various ways. We consider two models here, *viz.* (i) supersymmetry, (ii) little Higgs model, the first one very briefly and the second in little more detail.

Supersymmetry

In *supersymmetry*, every type of boson has a corresponding type of fermions as *superpartners* and vice versa. This theory is able to alleviate the difficulty of the instability of the Higgs mass in SM. This is because the contributions from the loops due to a particle and its superpartner come with opposite signs and therefore the divergent loop contributions in the quantum corrections of the Higgs mass cancel out. For example, in Fig.2 we show schematically, the cancellation of contributions from a gauge boson loop and the corresponding *gaugino* loop. Similar cancellations will be possible for a fermion loop and the corresponding superpartner loop. In the minimally supersymmetric standard model (MSSM) [5], the wider spectrum of particles than in the SM require wider spectrum of Higgs, namely, scalar Higgs $(h, H)$, pseudoscalar Higgs $(A)$, charged Higgs $(H^{\pm})$.

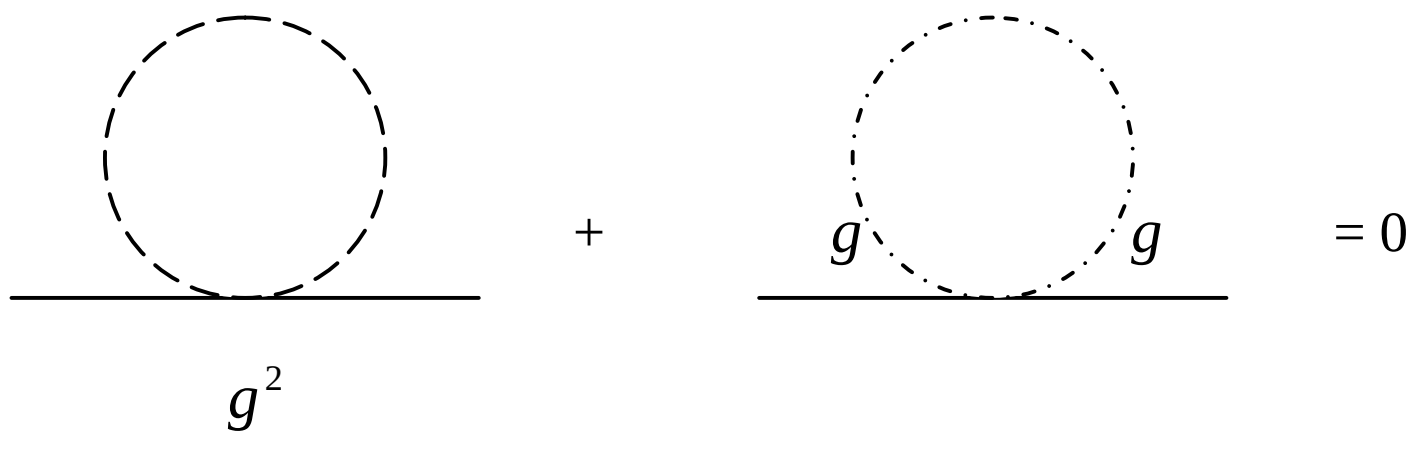

Fig.2. Cancellation of divergent contributions to the Higgs mass from a gauge boson loop (dashed line) and the gaugino (corresponding superpartner) loop (dash-dotted line).

<u>Little Higgs Model</u>

In the little Higgs model (LHM)[6], the Higgs particles appear as a subset of Goldstone bosons. There are various versions of the LHM. However, the littlest Higgs model ($L^2$HM)[7] is considered to be the most economical, which we describe here. The $L^2$HM has a global SU(5) symmetry which spontaneously breaks to SO(5) giving fourteen massless Goldstone bosons, $\eta, \omega^-, \omega^0, \omega^+, H^-, H^0, H^{0^*}, H^+, \phi^{--}, \phi^-, \phi^0, \phi_p^*, \phi^+$ and $\phi^{++}$. The gauging explicitly breaks the SU(5) symmetry at the TeV scale by a vacuum condensate which is proportional to,

$$\Sigma_0 = \begin{bmatrix} 0 & 0 & I_{2\times 2} \\ 0 & 1 & 0 \\ I_{2\times 2} & 0 & 0 \end{bmatrix}. \qquad (1)$$

By this symmetry breaking, massive gauge bosons occur by eating up $\eta, \omega^-, \omega^0$ and $\omega^+$ leaving a Higgs doublet,

$H = \begin{pmatrix} H^+ \\ H^0 \end{pmatrix}$ and a Higgs triplet, $\phi = \begin{pmatrix} -i\phi^{++} & -\frac{i\phi^+}{\sqrt{2}} \\ -\frac{i\phi^+}{\sqrt{2}} & \frac{-i\phi^0 + \phi_p^0}{\sqrt{2}} \end{pmatrix}$. (2)

So far as the gauge bosons in this model are concerned, there are two types: the heavy $SU(2)$ gauge bosons, $W^a_{H\mu}, a = 1, 2, 3$ which get mass by the explicit symmetry breaking (ESB) mentioned above and $W^a_{L\mu}, a = 1, 2, 3$ which get mass by an SSB, which is an electroweak symmetry breaking (EWSB). Similarly, there are two types of $U(1)$ gauge bosons: $B_H$ and $B_L$. The Higgs also gets mass by this SSB. The SSB can give rise to a vacuum expectation value (VEV) of the Higgs field equal to 246 GeV as in SM. However, it can give a higher value of the VEV, viz., 1.1 TeV corresponding to a vacuum condensate of the nonlinear sigma field as [8],

$$\Sigma = \begin{pmatrix} 0 & 0 & 0 & 1 & 0 \\ 0 & -s^2 & i\sqrt{2}sc & 0 & 1-s^2 \\ 0 & i\sqrt{2}sc & 1-2s^2 & 0 & i\sqrt{2}sc \\ 1 & 0 & 0 & 0 & 0 \\ 0 & 1-s^2 & i\sqrt{2}sc & 0 & -s^2 \end{pmatrix}, \qquad (3)$$

where, $s(c) = \sin\alpha(\cos\alpha)$, $\alpha = h/\sqrt{2}f$, $h$ being the physical Higgs field and $f$ is a scale. As has been stated, there are two sets of gauge fields. These two sets together participate in the SSB. Similarly, two sets of fermions (here, top quarks) participate in the SSB. Therefore, SSB in this case is called *collective symmetry breaking*. This collective symmetry breaking stabilizes the Higgs mass by turning the quadratic divergence into a logarithmic divergence. Without going into the detailed algebra we just point out that cancellations occur between the contributions of the loops of heavy and light particles, as shown in the following figures.

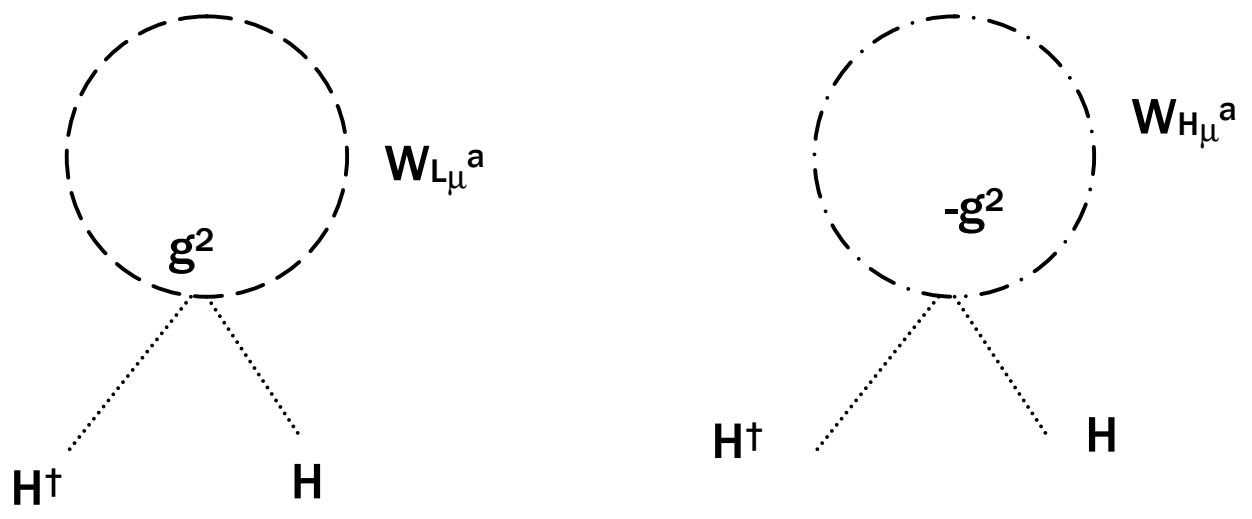

Fig.3 The quadratically divergent gauge-loop diagrams in the $L^2$HM which cancel out by collective symmetry breaking.

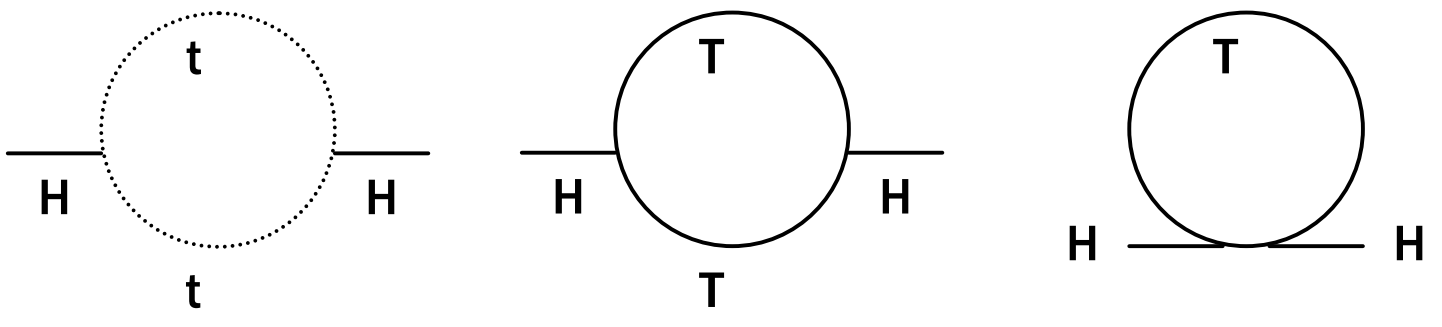

Fig.4 The quadratically divergent top quark-loop diagrams in the $L^2$HM which cancel out by collective symmetry breaking, subject to a relation between the coupling constants. Here, t is the light top quark and T the heavy top quark.

## 4. The role of Higgs in the early Universe

The symmetry breakings mentioned earlier are believed to have taken place in the early Universe at $t \sim 10^{-13}$ sec to $10^{-11}$ sec after the big bang. Thus, Higgs has important role to play in the evolution of the early Universe. The EWSB is usually associated with an electroweak phase transition (EWPT). The variation of the Higgs potential with temperature decides the order of the EWPT. It has been observed that a strong first-order EWPT provides the necessary mechanism for the generation of matter-antimatter asymmetry in the early Universe. In a kind of $L^2$HM, we get such an EWPT [8] and thereby an explanation [9] of this asymmetry.

## 5. Higgs search plans

The Higgs will be searched in the Large Hadron Collider (LHC) experiments by the CMS (Compact Muon Solenoid) and the ATLAS (A Toroidal LHC Apparatus) detectors, through the processes shown in Fig.5. Among the production channels depicted below, the gluon-gluon fusion has the highest cross-sections, viz., 10 pb at $m_H = 400$ GeV.

Close to LEP limits, the dominant decay modes of Higgs are, $H \rightarrow \gamma\gamma, \tau^+\tau^-, b\bar{b}$ and for the higher masses, the decay modes are, $H \rightarrow W^+W^-, ZZ$ [11]. The $b\bar{b}$ mode is associated with a large background. In that respect, the $\gamma\gamma$ decay is cleaner; however, it has got a small branching ratio. The most promising Higgs signal in the range $m_H = 140\text{-}600$ GeV is expected to come from, $H \rightarrow ZZ \rightarrow l^+l^-l^+l^-$, which can be made practically background-free. For $m_H =$600-1000 GeV, the favourable signals will come from, $H \rightarrow WW \rightarrow l\nu q\bar{q}$ and $H \rightarrow ZZ \rightarrow l^+l^-\nu\nu$.

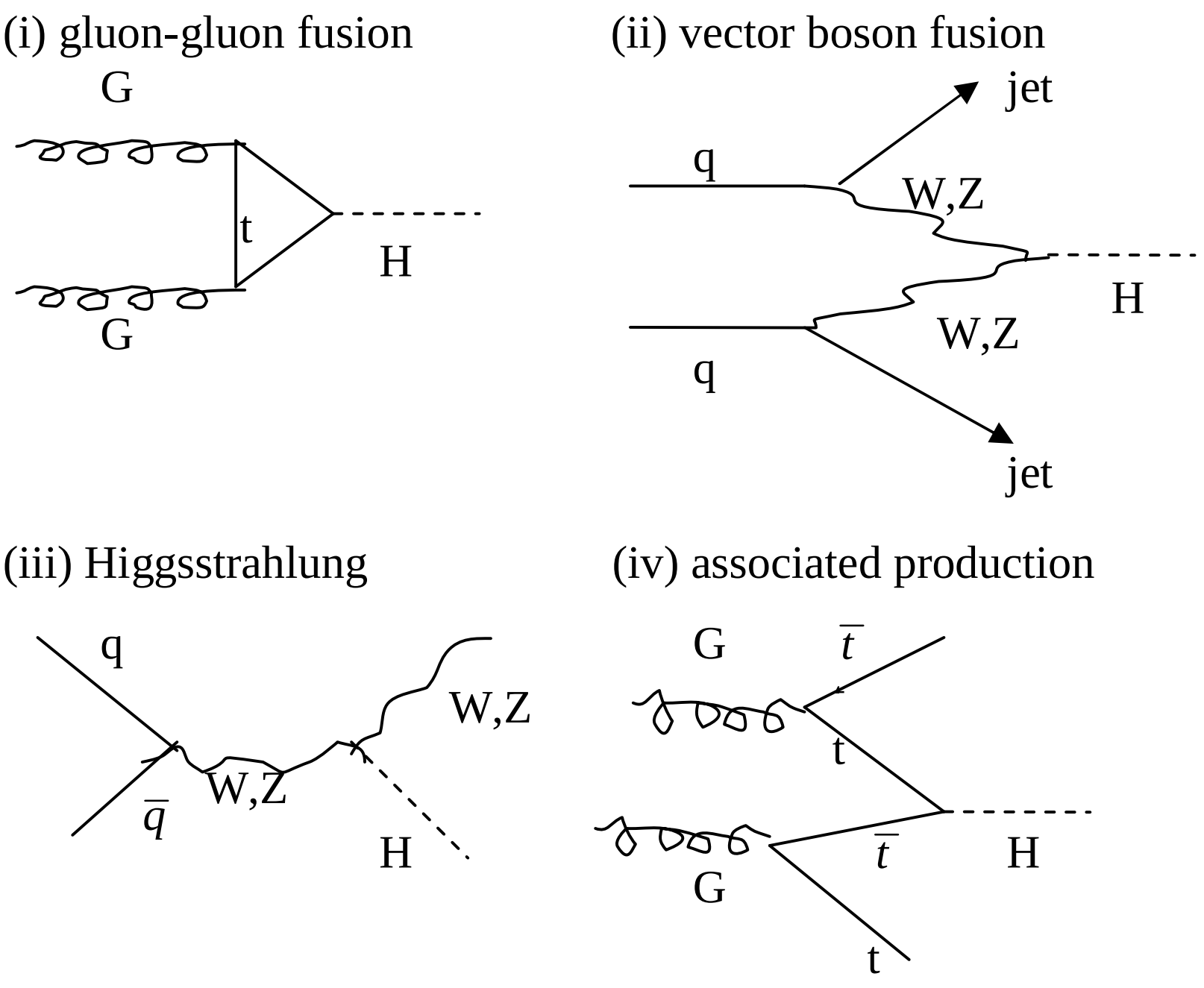

Fig.5 The SM Higgs production channels at the LHC. Here G is a gluon and q a quark.

## 6. Conclusions

The Higgs boson is being intensely searched both at the Tevatron as well as the LHC. The present status of such searches can be obtained in Ref.12.